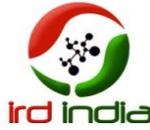

# Document Clustering using K-Medoids

Monica Jha

Department of Information and Technology, Gauhati University, Guwahati, India
Email: monicajha88@gmail.com

**Abstract-** People are always in search of matters for which they are prone to use internet, but again it has huge assemblage of data due to which it becomes difficult for the reader to get the most accurate data. To make it easier for people to gather accurate data, similar information has to be clustered at one place. There are many algorithms used for clustering of relevant information in one platform. In this paper, K-Medoids clustering algorithm has been employed for formation of clusters which is further used for document summarization.

**Keywords:** Clustering, Document Summarization, K-medoids, Matrix

## I. INTRODUCTION

Gathering up the most relevant data to one's need, from the huge collection of data in the internet is a work of great hazard. To make things easier, document clustering was introduced. Document clustering is a method of arranging similar type of data into a cluster which would differ from the data of another cluster. Again, there are different ways of clustering based on the feature selected to be used. In this paper, we will see how clustering is achieved using K-Medoids clustering algorithm, which later is used for document summarization based on the highest weighed sentence.

The organization of the paper is as follows. Section II describes the related work on topic. In section III the design and implementation of the system has been discussed. Section IV gives a description on K-MEDOIDS. In section V gives a review on text-document summarization. Results obtained from our experiment are showed on section VI. Section VII concludes the paper and section VIII discusses the limitations of the existing systems.

## II. REVIEW

Document clustering has been developed in order to make it easier for us when we go for information retrieval. It helps in retrieving most accurate collection of documents whenever say we type in on net to get information on a particular topic. Till now many different methods has been applied in document clustering. In the paper [1], the author proposed a method which aims to cluster the documents into different semantic classes. how it is difficult to employ clustering in a proper way because of its high dimensional space which is unworkable due to the curse of dimensionality. So, they proposed Locality Preserving Indexing into lower- dimensional semantic space. In the paper [2], authors Levent Bolelli, Seyda Ertekin, Ding Zhou and C. Lee Giles has laid the emphasize on K-SVMeans, a clustering algorithm for multi-type interrelated datasets that integrates the well-known K-Means clustering with the highly popular Support Vector Machines. According to them, their experiment shows that K-SVMeans yields better result than homogeneous data clustering. Mehrdad Mahdavi and Hassan Abolhassani in their paper [3] has laid stress on how they used Harmony K-means Algorithm that deals with document clustering based on Harmony Search optimization method. In their paper they reveal that their experimental results that Harmony K-means Algorithm converges to the best known optimum faster than other methods and the quality of clusters are comparable. In the paper [4], authors Maria Camila N. Barioni, Humberto L. Razente, Agma J. M. Traina and Caetano Traina Jr based on several datasets, including synthetic and real data, show that the proposed algorithm may reduce the number of distance calculations by a factor of more than a thousand times when compared to existing algorithms while producing clusters of comparable quality. Michael Steinbach, George Karypis and Vipin Kumar in their paper [5] have discussed the comparison between agglomerative hierarchical clustering and K-means. In the paper they have shown how bisecting K-means technique is better than the standard K-means approach and as good as or better than the hierarchical approaches that they tested for a variety of cluster evaluation metrics. Further in the paper they propose an explanation based on these algorithms. The paper [6] by Julian Sedding and Dimitar Kazakov tells us about how text clustering can simplify large collection of document clustering by dividing it into smaller clusters. In their research paper they have laid stress on primitive, syntax-based clarification by allotting each word a part-of-speech tag and by improving the data representation often used for document clustering with synonyms and hyponyms i.e. subordinate from Word Net. In the paper [7], authors have proposed an algorithm for documents that uses a sampling-based pruning strategy to simplify hierarchical clustering. According to them the algorithm that they have introduced can be used any text document data set whose entries can be embedded in a high dimensional Euclidean space in which every document is a vector of real numbers.

## III. METHODOLOGY

An algorithms output usually depends on the input provided to it. The input provided to the algorithm, were polished following some steps. All the steps were carried out in Java IDE. Randomly selected documents from different domains (Entertainment, Literature, Sport, Music, Technology and Political) were collected in the beginning.

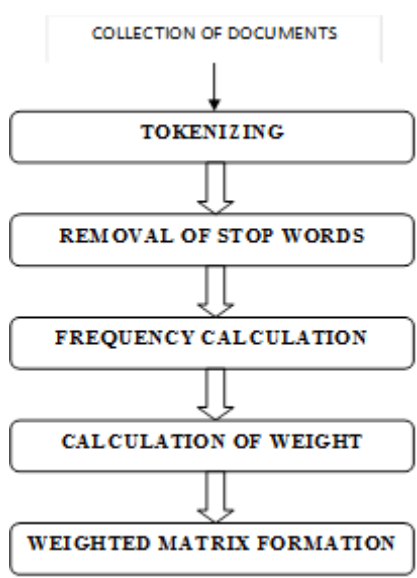

Fig1. Steps in document clustering

Example:

Entertainment: Salaam Bombay is a 1988 Hindi film directed by Mira Nair, and screen written by her longtime creative collaborator, Sooni Taraporevala. The film chronicles the day-to-day life of children living on the streets of Mumbai. It won the National Film Award for Best Feature Film in Hindi, the National Board of Review Award for Top Foreign Film, the Golden Camera and Audience Awards at the Cannes Film Festival, and three awards at the Montréal World Film Festival. The film was India's second film submission to be nominated for the Academy Award for Best Foreign Language Film.

Literature: In the 12th century, a new form of English now known as Middle English evolved. This is the earliest form of English literature which is comprehensible to modern readers and listeners, albeit not easily. Middle English lasts up until the 1470s, when the Chancery Standard, a form of London-based English, became widespread and the printing press regularized the language. Middle English Bible translations, notably Wycliffe's Bible, helped to establish English as a literary language.

Sport: Badminton is a racquet sport played by either two opposing players (singles) or two opposing pairs (doubles), who take positions on opposite halves of a rectangular court that is divided by a net. Players score points by striking a shuttlecock with their racquet so that it passes over the net and lands in their opponents' half of the court.

Literature: The Bharatiya Janata Party (abbreviated BJP) is one of the two major parties in the Indian political system, the other being the Indian National Congress. Established in 1980, it is India's second largest political party in terms of representation in the parliament and in the various state assemblies. The Bharatiya Janata Party designates its official ideology and central philosophy to be "integral humanism", based upon a 1965 book by Deendayal Upadhyaya. The party advocates Hindu nationalism and social conservatism, self-reliance as outlined by the Swadeshi movement and a foreign policy centered around key nationalist principles.

Zoology: Evolution is the change in the inherited characteristics of biological populations over successive generations. Evolutionary processes give rise to diversity at every level of biological organization, including species, individual organisms and molecules such as DNA and proteins. Life on Earth evolved from a universal common ancestor approximately 3.8 billion years ago.

Similarly, documents related to various topics of the different domain were taken into account in the form of text documents.

Tokenizing: The documents are assembled in a folder after which we assemble each and every word of all the hundred documents in a single file and then we tokenize that file i.e., break the stream of texts up into words and list them one by one.

Removal of Stopwords: The next step involved removal of those words which are used in huge numbers but their usage can lead to the degradation of the performance therefore they are removed. These words are termed as stopwords.

Frequency Calculation: In this step, the frequency of each word of the dictionary is calculated. This is done by checking in the occurrence of each word in the dictionary and also the total word of the document is checked. Frequency map was used in order to calculate term frequency in the program.

Weight Calculation: The next step involves calculating the weight of each word. This is calculated by taking the ratio of frequency of each word to the total number of words in the document.

WEIGHT = FREQUENCY OF WORD/TOTAL No. OF WORDS IN THE DOCUMENT

**Vector Space Model:**

Documents are represented as vectors in term space. Vector space model helps in performing two very important steps:

1) Removal of stop-words as they are frequent and carry no information. Stop words example- a, an, the, how, when, etc.

2) Stemming of the words to its original form. For example, the stem word of defined is define.

The Vector Space Model is the basic model for document clustering, upon which many modified models are based. In this model, each document Dj is first represented as a term frequency vector in the term-space:

Djtf = (tf1j; tf2j : : : : : : : ; tfVj)0 j = 1; 2; ::::;D

Where tfij is the frequency of the i-th term in document dj, V is the total number of the selected vocabulary, and D is the total number of documents in the collection.

Weight formula that has been employed in the project is given below:

Weight = tf * idf

Where tf is term frequency and idf is inverse document frequency.

$$\begin{pmatrix} & T_1 & T_2 & \cdots & T_t \\ D_1 & w_{11} & w_{21} & \cdots & w_{t1} \\ D_2 & w_{12} & w_{22} & \cdots & w_{t2} \\ \vdots & \vdots & \vdots & & \vdots \\ \vdots & \vdots & \vdots & & \vdots \\ D_n & w_{1n} & w_{2n} & \cdots & w_{tn} \end{pmatrix}$$

Fig 2. Matrix

**Matrix Formation**

In a matrix, weight of each word in rows haws stored and in the column, the document name was stored. When work on K-Medoids clustering algorithm begins, this matrix is taken as the input of the algorithm and then further implementation of the algorithm is done on this matrix. The matrix is given in the fig 2.

## IV. K-MEDOIDS

As discussed earlier K-medoids is a type of partition algorithm. This algorithm is somewhat similar to k-means algorithm which is again another type of partition algorithm. The only extra restraint that k-medoids add is that the centers that are used to represent the data are taken from the dataset itself. Therefore, medoid is a data-point that represents a set of data set. This algorithm is used when we have a huge data set. This was introduced in order to overcome the limitations of k-means algorithm.

In the next few steps we will see the different steps followed in the algorithm:

**Cluster of documents of various domains**

This is the first step of the algorithm. Here, we are supposed to arbitrarily choose medoids from the data set.

Step1. Medoids Initialization

This is the first step of the algorithm. Here, arbitrarily medoids are selected from the data set. Five is selected as medoid number and each one of the medoid is from five different genres which have been mentioned above.

Step2. Manhattan distance calculation

The next step is to calculate the Manhattan distance with each of the documents from the respective selected medoids.

According to Manhattan distance,

$d(i, j) = |X_{i1}-X_{j1}| + |X_{i2}-X_{j2}| + ... + |X_{in} -X_{jn}|$

All the rest of the documents which has not been selected as the medoids is then employed to find in the manhattan distances from each of the selected medoids respectively.

Here, in the above formula,

i implies the document id of each of the medoid respectively which keeps on changing after one gets the manhattan distance with each of the non selected documents.

j implies the document id of each of the non- selected documents which changes to other non-selected document after the manhattan distance of it is found out with each of the selected medoids.

Step 3: Cluster formation

In this step, firstly distances of each of the documents with each of the medoids is determined. And then a comparison of each of the obtained medoids is performed to find in the least manhattan distance which is considered to be the best one. Hence, that document forms cluster with the medoid with which least manhattan distance was obtained. Similarly, all the documents distance is checked with each of the selected medoids respectively to find in the least distance to form in the cluster. Therefore, the clusters are derived. Distance calculation could have been calculated using Euclidean distance as well but we selected manhattan distance for it can used with much ease.

The cluster formation in the project is based on the distances of the documents with the medoids which is basically based on the similarity of words in the same documents and different in the other clusters. Hence, five different clusters are formed based on the medoid selection of different genre.

Step 4. Calculating total cluster distance

The next step involves finding out the total distance of the each of the cluster formed.

Step 5. Selection of non-medoid object

The next step involves selection of a non-medoid object randomly and replaces it with one of the existing medoid. After which we are suppose to again use the Manhattan distance to find the distance with each and every weights of each corresponding documents.

Step 6. Comparison of distance with new and old medoids

The next step involves finding out the Manhattan distance with the new medoids. After which the cluster is formed by comparing the distance achieved of both the medoids. Lastly, the total distance of the new clusters is calculated and in the end, the total distance of the old cluster with the new cluster formed is compared; whichever is the minimum one is retained.

Step 7. Repeat until no change

The above steps are repeated until medoids which have lesser distance as compared to the prior or later selected medoids is achieved.

## V. MULTI-DOCUMENT SUMMARIZATION

When the clusters are formed the next step is that of document summarization. Document summarization is carried out as it provides meaningful summary for each document. This enables the user to judge the documents in a proper way and to read only those documents which are relevant to their need. Document summarization can be carried out for a single document as well as for a number of documents at a time. The desired summary can be obtained in many forms: on paragraph basis, keyword basis etc.

Generally there are two different approaches for document summarization: Abstractive approach and Extractive approach.

**Abstractive approach:** in this approach summary is generated in abstract form. One can obtain useful and well-defined domains with clear cut information.

**Extractive approach**: this approach yields sentences or clauses. One can obtain arbitrarily complex and unstructured domains using this approach.

## VI. RESULTS

Documents from five different domain was taken into consideration(total hundred documents, twenty from each domain), which was put in refining and then the result obtained in the form of matrix, which contained the weight of terms was employed for cluster formation using K-Medoids Algorithm.

| CLUSTER No | No of DOCUMENTS | EFFICIENCY |
|---|---|---|
| 0 | 43 | 18.60% |
| 1 | 11 | 54.15% |
| 2 | 7 | 42.85% |
| 3 | 35 | 14.20% |
| 4 | 4 | 50.00% |

TABLE I. OBSERVATION TABLE

In the table I, we can see the number of documents in each cluster along with the efficiency. The formed clustered are then processed further for document summarization which is obtained by key points of each documents based on weight of each sentence of a document.

## VII. CONCLUSION

After implementing the algorithm, it was observed with the increase in number of documents, the result obtained is better one. The result calculated using term frequency in case of K-Medoids algorithm is better than using tf-idf. In the end, after clusters are formed its shown how those clusters are further used for summarization, which show cases the highest weighed sentences of each of the document of the best clusters formed. This project helps in providing the input for document summarization hence, making it easier for the researchers to save time and also obtain result with much ease.

## VIII. LIMITATIONS

The main limitation of the project was when a cluster is formed; there is always a huge possibility that there might be a few other domain documents in the cluster of other domain documents due to the similar number of words in those documents. Moreover the time required for calculation of term frequency-inverse document frequency weight is quite high, it consumes a lot of time during calculation of the algorithm.

The main focus of the project was to build an input for document summarization. That work has been implemented in the project using K-Medoids clustering based algorithm using two different inputs for the algorithm using two different datasets. The datasets were compared to find that the greater the number of the documents the better the results. In future, the datasets can be increased to obtain yet better results which could be used for the processing of better document summarization.